\documentclass[12pt]{article}

\usepackage{graphicx}
\begin{document}

\begin{center}
{\bf Dyonic and magnetic black holes with rational nonlinear electrodynamics} \\
\vspace{5mm} S. I. Kruglov
\footnote{E-mail: serguei.krouglov@utoronto.ca}
\underline{}
\vspace{3mm}

\textit{Department of Physics, University of Toronto, \\60 St. Georges St.,
Toronto, ON M5S 1A7, Canada\\
Canadian Quantum Research Center,\\ 204-3002 32 Ave Vernon, BC V1T 2L7, Canada} \\
\vspace{5mm}
\end{center}

\begin{abstract}
The principles of causality and unitarity are studied within rational nonlinear electrodynamics proposed earlier. We investigate dyonic and magnetized black holes and show that in the self-dual case, when the electric charge equals the magnetic charge, corrections to Coulomb's law and Reissner$-$Nordstr\"{o}m solutions are absent. In the case of the magnetic black hole, the Hawking temperature, the heat capacity and the Helmholtz free energy are calculated. It is shown that there are second-order phase transitions and it was demonstrated that at some range of parameters the black holes are stable.
\end{abstract}

\section{Introduction}

The black holes (BHs) are real objects in the centers of many galactic and its physics is of great interest.   Dyonic solutions in the string \cite{Shapere}-\cite{Jatkar} and in the supergravity \cite{Chamseddine}-\cite{Pang} theories for BHs with magnetic and electric charges were obtained. Such solutions are used in the theory of superconductivity and thermodynamics \cite{Hartnoll}, \cite{Hartnoll1}, \cite{Dutta}.
In this paper we obtain dyonic and magnetic BH solutions in the framework of rational nonlinear electrodynamics proposed in \cite{Krug2}. The attractive feature of this nonlinear electrodynamics (NED) is the absence of singularities in the center of charges and their finite self-energy. Similar properties of NED were firstly observed by Born and Infeld in another NED \cite{Born}. Quantum electrodynamics with loop corrections also leads to NED \cite{Heisenberg}. The singularity problems are absent also in other NED models \cite{Soleng}-\cite{Kruglov1}. The general relativity (GR) and the thermodynamics of BH with NED was considered in \cite{Pellicer}-\cite{Krug6}. The phase transitions in electrically and magnetically charged BHs were investigated in \cite{Yajima}-\cite{Kruglov3}. It worth noting that the universe acceleration also can be explained by NED coupled with GR \cite{Garcia}-\cite{Kruglov4}.

The paper is organised as follows. In Sec. 2 we study the causality and unitarity principles. We obtain the dyonic solution in Sec. 3. In Sec. 4 we consider the magnetic BH. The metric function and their asymptotic as $r\rightarrow\infty$  are found. It was shown that the magnetic mass of BHs is finite and there are not singularities of the Ricci scalar as $r\rightarrow\infty$. The BH thermodynamics and the thermal stability of charged black holes are investigated in Sec. 5. We obtain the Hawking temperature, the heat capacity, the Helmholtz free energy and demonstrate that the phase transitions in BHs occur. 

We use units with $c=1$ and the metric signature $\mbox{diag}(-1,1,1,1)$.

\section{The model and principles of causality and unitarity}

Here, we consider rational NED, proposed in \cite{Krug2}, with the Lagrangian density
\begin{equation}
{\cal L} = -\frac{{\cal F}}{2\beta{\cal F}+1},
 \label{1}
\end{equation}
where the parameter $\beta\geq 0$ possesses the dimension of (length)$^4$, ${\cal F}=(1/4)F_{\mu\nu}F^{\mu\nu}=(B^2-E^2)/2$, $F_{\mu\nu}=\partial_\mu A_\nu-\partial_\nu A_\mu$ is the field tensor.
The symmetrical energy-momentum tensor is given by \cite{Krug4}
\begin{equation}
T_{\mu\nu}=-\frac{F_\mu^{~\alpha}F_{\nu\alpha}}{(1+2\beta{\cal F})^{2}}
-g_{\mu\nu}{\cal L}.
\label{2}
\end{equation}
From Eq. (2) we obtain the energy density
\begin{equation}
\rho=T_0^{~0}=\frac{{\cal F}}{1+2\beta{\cal F}}
+\frac{E^2}{(1+2\beta{\cal F})^2}.
\label{3}
\end{equation}

For healthy theory the general principles of causality and unitarity should hold. According to the causality principle the group velocity of excitations over the background has to be less than the light speed, and then  tachyons are absent in the theory. The absence of ghosts is guaranteed by the unitarity principle. Both principles are satisfied for the case $\textbf{E}\cdot\textbf{B}=0$  if the following inequalities hold \cite{Shabad2}:
\[
 {\cal L}_{\cal F}\leq 0,~~~~{\cal L}_{{\cal F}{\cal F}}\geq 0,
\] \begin{equation}
{\cal L}_{\cal F}+2{\cal F} {\cal L}_{{\cal F}{\cal F}}\leq 0,
\label{4}
\end{equation}
where ${\cal L}_{\cal F}\equiv\partial{\cal L}/\partial{\cal F}$.
Making use of Eq. (1) we obtain
\[
{\cal L}_{\cal F}= -\frac{1}{(1+2\beta{\cal F})^2},
\]
\begin{equation}
{\cal L}_{\cal F}+2{\cal F} {\cal L}_{{\cal F}{\cal F}}=\frac{6\beta{\cal F}-1}{(1+2\beta{\cal F})^3},~~~~
{\cal L}_{{\cal F}{\cal F}}=\frac{4\beta}{(1+2\beta{\cal F})^3}.
\label{5}
\end{equation}
With the help of Eqs. (4) and (5), the principles of causality and unitarity take place if $6\beta{\cal F}\leq 1$ ($\beta\geq 0$). When $\textbf{E}=0$, $\beta\textbf{B}^2\leq 1/3$.

\section{The dyonic solution}

The action of NED coupled with GR is given by
\begin{equation}
I=\int d^4x\sqrt{-g}\left(\frac{1}{16\pi G}R+ {\cal L}\right),
\label{6}
\end{equation}
where $G$ is Newton's constant, $16\pi G\equiv M_{Pl}^{-2}$, and $M_{Pl}$ is the reduced Planck mass. The Einstein equation is
\begin{equation}
R_{\mu\nu}-\frac{1}{2}g_{\mu\nu}R=-8\pi GT_{\mu\nu}.
\label{7}
\end{equation}
Varying action (6) on electromagnetic potentials we obtain the fields equation for electromagnet fields
\begin{equation}
\partial_\mu\left(\sqrt{-g}F^{\mu\nu}{\cal L}_{\cal F}\right)=0.
\label{8}
\end{equation}
We consider the static and spherically symmetric metric with the line element
\begin{equation}
ds^2=-A(r)dt^2+\frac{1}{A(r)}dr^2+r^2(d\vartheta^2+\sin^2\vartheta d\phi^2),
\label{9}
\end{equation}
where the metric function is given by
\begin{equation}
A(r)=1-\frac{2M(r)G}{r},
\label{10}
\end{equation}
and the mass function is
\begin{equation}
M(r)=m_0+\int^r_0\rho(r)r^2dr=m_0+m_{el}-\int^\infty_r\rho(r)r^2dr.
\label{11}
\end{equation}
The total mass of the BH $m=m_0+m_{el}$, where $m_0$ is the Schwarzschild mass and $m_{el}=\int^\infty_0\rho(r)r^2dr$  is the electromagnetic mass.
The general solutions of field equations, found in \cite{Bronnikov3}, \cite{Bronnikov4}, are given by
\begin{equation}
B^2=\frac{q^2_m}{r^4},~~~~E^2=\frac{q_e^2}{{\cal L}^2_{\cal F}r^4},
\label{12}
\end{equation}
where $q_m$ and $q_e$ are the magnetic and electric charges, respectively.
With the help of Eqs. (1) and (12) one finds
\begin{equation}
E^2=\frac{q_e^2(1+2\beta{\cal F})^4}{r^4},
\label{13}
\end{equation}
\begin{equation}
\beta{\cal F}=a-b(1+2\beta{\cal F})^4,~~~a=\frac{\beta q^2_m}{2r^4},~~~
b=\frac{\beta q_e^2}{2r^4},
\label{14}
\end{equation}
and we introduced the unitless variables $a$ and $b$. Defining the unitless value $x\equiv\beta{\cal F}$, we obtain from Eq. (14) the equation as follows:
\begin{equation}
b(2x+1)^4+x-a=0.
\label{15}
\end{equation}
Using unitless variables $t=r/\sqrt[4]{\beta q_m^2}$ and $n=q_m^2/q_e^2$, one finds from Eq. (15) the equation for $y=2x+1$:
\begin{equation}
y^4+t^4y-n-t^4=0.
\label{16}
\end{equation}
The real dyonic solution to Eq. (16) is
\[
y=\sqrt{\frac{\sqrt[4]{3}t^4}{4\sqrt[4]{n+t^4}\sqrt{\sinh(\varphi/3)}}-\frac{\sqrt{n+t^4}\sinh(\varphi/3)}
{\sqrt{3}}}-\frac{\sqrt{\sinh(\varphi/3)}\sqrt[4]{n+t^4}}{\sqrt[4]{3}},
\]
\begin{equation}
\sinh(\varphi)=\frac{3^{3/2}t^8}{16(n+t^4)^{3/2}}.
\label{17}
\end{equation}
Putting $n=0$ in Eq. (17) we come to the solution corresponding to the electrically charged BH \cite{Krug4}. We find the self-dual solution at $q_e=q_m$ ($a=b$) from Eq. (15). Then $x=0$ (${\cal F}=0$, $E=B$), $E=q/r^2$ ($q\equiv q_e=q_m$) and with the help of Eqs. (3) and (11) we obtain the mass function
\begin{equation}
M(r)=m-\int^\infty_r\rho(r)r^2dr=m-\frac{q^2}{r}.
\label{18}
\end{equation}
Making use of Eq. (10) one finds the metric function
\begin{equation}
A(r)=1-\frac{2mG}{r}+\frac{2q^2G}{r^2}.
\label{19}
\end{equation}
The metric function (19) corresponds to the Reissner$-$Nordst\"{o}m (RN) solution with $2q^2=q^2_e+q^2_m$.

\section{The magnetic black hole}

Let us consider the static magnetic BH \footnote{In the paper M.-S. Ma, Ann. Phys. \textbf{362}, 529 (2015) the author also considered the static magnetic BH based on NED proposed in \cite{Krug2}. However, here we use unitless variables that are more convenient for the analyses of the BH thermodynamics. In addition, we analyse more general case when the BH besides the electromagnetic mass possesses the Schwarzschild mass (having non-electromagnetic nature).}. Taking into account that $q_e=0$, ${\cal F}=q_m^2/(2r^4)$, we obtain from Eq. (3) the magnetic energy density
\begin{equation}
\rho_M=\frac{B^2}{2(\beta B^2+1)}=\frac{q_m^2}{2(r^4+\beta q_m^2)}.
\label{20}
\end{equation}
With the help of Eqs. (11) and (20) one finds the mass function
\[
M(x)=m_0+\frac{q_m^{3/2}}{8\sqrt{2}\beta^{1/4}}\biggl(\ln\frac{x^2-\sqrt{2}x+1}{x^2+\sqrt{2}x+1}
\]
\begin{equation}
+2\arctan(\sqrt{2}x+1)-2\arctan(1-\sqrt{2}x)\biggr),
\label{21}
\end{equation}
where $x=r/\sqrt[4]{\beta q_m^2}$.
The BH magnetic mass is given by
\begin{equation}
m_M=\int_0^\infty\rho_M(r)r^2dr=\frac{\pi q_m^{3/2}}{4\sqrt{2}\beta^{1/4}}\approx 0.56\frac{q_m^{3/2}}{\beta^{1/4}}.
\label{22}
\end{equation}
The Schwarzschild mass $m_0$ is a free parameter and at $q_m=0$ one has $m_M=0$, and we arrive at the Schwarzschild BH. Making use of Eq. (10) we obtain the metric function
\[
A(x)=1-\frac{2m_0G}{\sqrt[4]{\beta q_m^2}x}-\frac{q_mG}{4\sqrt{2\beta}x}\biggl(\ln\frac{x^2-\sqrt{2}x+1}{x^2+\sqrt{2}x+1}
\]
\begin{equation}
+2\arctan(\sqrt{2}x+1)-2\arctan(1-\sqrt{2}x)\biggr),
\label{23}
\end{equation}
As $r\rightarrow \infty$ the metric function (23) becomes
\begin{equation}
A(r)=1-\frac{2mG}{r}+\frac{q_m^2G}{r^2}+{\cal O}(r^{-5})~~~~r\rightarrow \infty,
\label{24}
\end{equation}
where $m=m_0+m_M$.
The correction to the RN solution, according to Eq. (24), is in the order of ${\cal O}(r^{-5})$.
At $m_0=0$ and $r\rightarrow 0$, from Eq. (23) one finds the asymptotic with a de Sitter core
\begin{equation}
A(r)=1-\frac{Gr^2}{\beta}+\frac{Gr^6}{7\beta^2q_m^2}-\frac{Gr^{10}}{11\beta^3q_m^4}+{\cal O}(r^{12})~~~~r\rightarrow 0.
\label{25}
\end{equation}
The solution (25) is regular because as $r\rightarrow 0$ we have $A(r)\rightarrow 1$. When $m_0\neq 0$ the solution is singular and $A(r)\rightarrow \infty$.
Let us introduce unitless constants $C=m_0G/(\beta^{1/4}\sqrt{q_m})$, $B=q_mG/\sqrt{\beta}$.
Then the horizon radii, that are the roots of the equation $A(r)=0$ ($x_{+/-}=r_{+/-}/(\sqrt{q_m}\beta^{1/4})$),  are given in Tables 1 and 2.
\begin{table}[ht]
\caption{$B=1$}
\centering
\begin{tabular}{c c c c c c c c c c  c}\\[1ex]
\hline
$C$ & 0.6 & 0.7 & 0.8 & 0.9 & 1 & 2 & 3 & 4 & 5 \\[0.5ex]
\hline
 $x_+$ & 1.75 & 2.02 & 2.27 & 2.52 & 2.75 & 4.91 & 6.97 & 9.00 & 11.02 \\[0.5ex]
\hline
\end{tabular}
\end{table}
\begin{table}[ht]
\caption{$m_0=0$}
\centering
\begin{tabular}{c c c c c c c c c c  c}\\[1ex]
\hline
$B$ & 3.173 & 3.2 & 3.5 & 4 & 4.5 & 5 & 6 & 7 & 8 \\[0.5ex]
\hline
 $x_-$ & 1.68 & 1.52 & 1.21 & 1.03 & 0.92 & 0.85 & 0.75 & 0.68 & 0.63 \\[0.5ex]
\hline
 $x_+$ & 1.68 & 1.87 & 2.49 & 3.19 & 3.82 & 4.42 & 5.59 & 6.74 & 7.87 \\[0.5ex]
\hline
\end{tabular}
\end{table}
The plots of the metric function (23) are depicted in Figs. 1 and 2.
\begin{figure}[h]
\centering
\includegraphics [height=2.in,width=2.in]{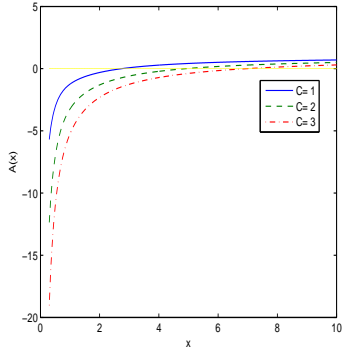}
\caption{\label{fig.1} The plot of the function $A(x)$ for $B=1$. The solid curve is for $C=1$, the dashed curve corresponds to $C=2$, and the dashed-doted curve corresponds to $C=3$.}
\end{figure}
\begin{figure}[h]
\centering
\includegraphics[height=2.in,width=2.in]{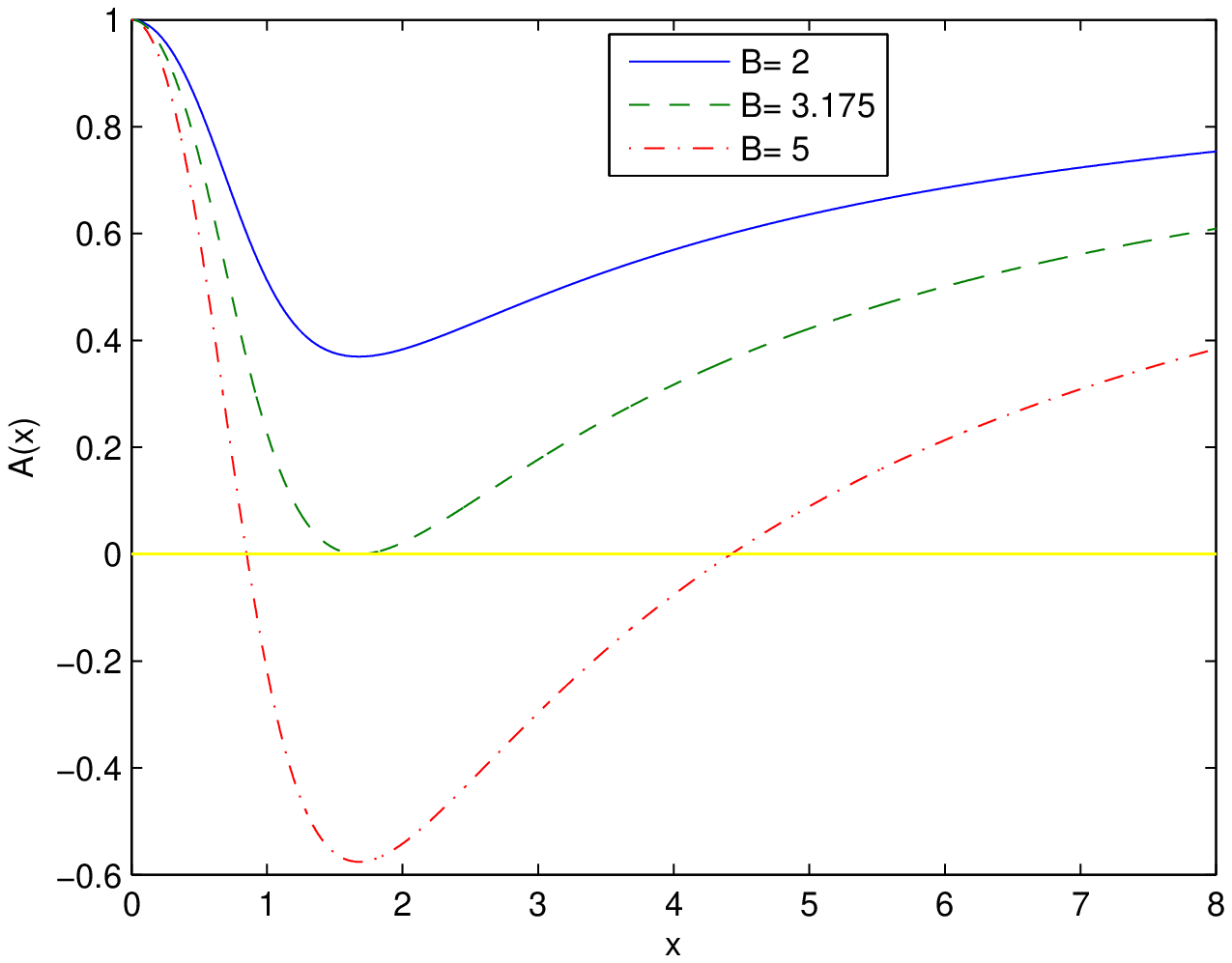}
\caption{\label{fig.2} The plot of the function $A(x)$ for $m_0=0$. The solid curve is for $B=2$, the dashed curve corresponds to $B=3.175$, and the dashed-doted curve corresponds to $B=5$.}
\end{figure}
According to Fig. 1 at $m_0\neq 0$ ($B=1$) there is only one horizon. For the bigger mass (the parameter $C$ is greater) the horizon radius increases. Figure 2 shows that there are no horizons at $m_0=0$, $B<3.17$), an extreme horizon occurs at $m_0=0$, $B\approx 3.173$, and two horizons hold at $m_0=0$ and $B>3.173$.

Making use of Eqs. (2) and (7) at $E=0$, we obtain the Ricci scalar
\begin{equation}
R=8\pi G T_\mu^{~\mu}=\frac{16\pi G\beta q_m^4}{(r^4+\beta q_m^2)^2}.
\label{26}
\end{equation}
The Ricci scalar approaches to zero as $r\rightarrow \infty$ and spacetime becomes flat.

\section{The black hole thermodynamics}

To study the black holes thermodynamics and the thermal stability of magnetic BHs, we consider the Hawking temperature
\begin{equation}
T_H=\frac{\kappa}{2\pi}=\frac{A'(r_+)}{4\pi},
\label{27}
\end{equation}
where $\kappa$ is the surface gravity and $r_+$ is the event horizon radius. With the help of Eqs. (10) and (11) one finds the relations
\begin{equation}
A'(r)=\frac{2 GM(r)}{r^2}-\frac{2GM'(r)}{r},~~~M'(r)=r^2\rho,~~~M(r_+)=\frac{r_+}{2G}.
\label{28}
\end{equation}
Making use of Eqs. (3), (27) and (28) we obtain the Hawking temperature
\begin{equation}
T_H=\frac{1}{4\pi}\left(\frac{1}{r_+}-2Gr_+\rho(r_+)\right)
=\frac{1}{4\pi\beta^{1/4}\sqrt{q_m}}\left(\frac{1}{x_+}-\frac{Gq_mx_+}{\sqrt{\beta}(1+x_+^4)}\right).
\label{29}
\end{equation}
Using the equation $M(r_+)=r_+/(2G)$ and (21) we find
\[
\frac{Gq_m}{\sqrt{\beta}}=\frac{4\sqrt{2}(x_+-2C)}{D},
\]
\begin{equation}
D\equiv \ln\frac{x^2-\sqrt{2}x+1}{x^2+\sqrt{2}x+1}
-2\arctan(1-\sqrt{2}x)+2\arctan(1+\sqrt{2}x).
\label{30}
\end{equation}
Replacing Eq. (30) into Eq. (29) one obtains the Hawking temperature as follows:
\begin{equation}
T_H==\frac{1}{4\pi\beta^{1/4}\sqrt{q_m}}\biggl(\frac{1}{x_+}
-\frac{4\sqrt{2}(x_+-2C)x_+}{(1+x_+^4)D}\biggr).
\label{31}
\end{equation}
The plots of the functions $T_H(x_+)\sqrt{q_m}\beta^{1/4}$ are depicted in Figs. 3 and 4.
\begin{figure}[h]
\centering
\includegraphics[height=2.in,width=2.in]{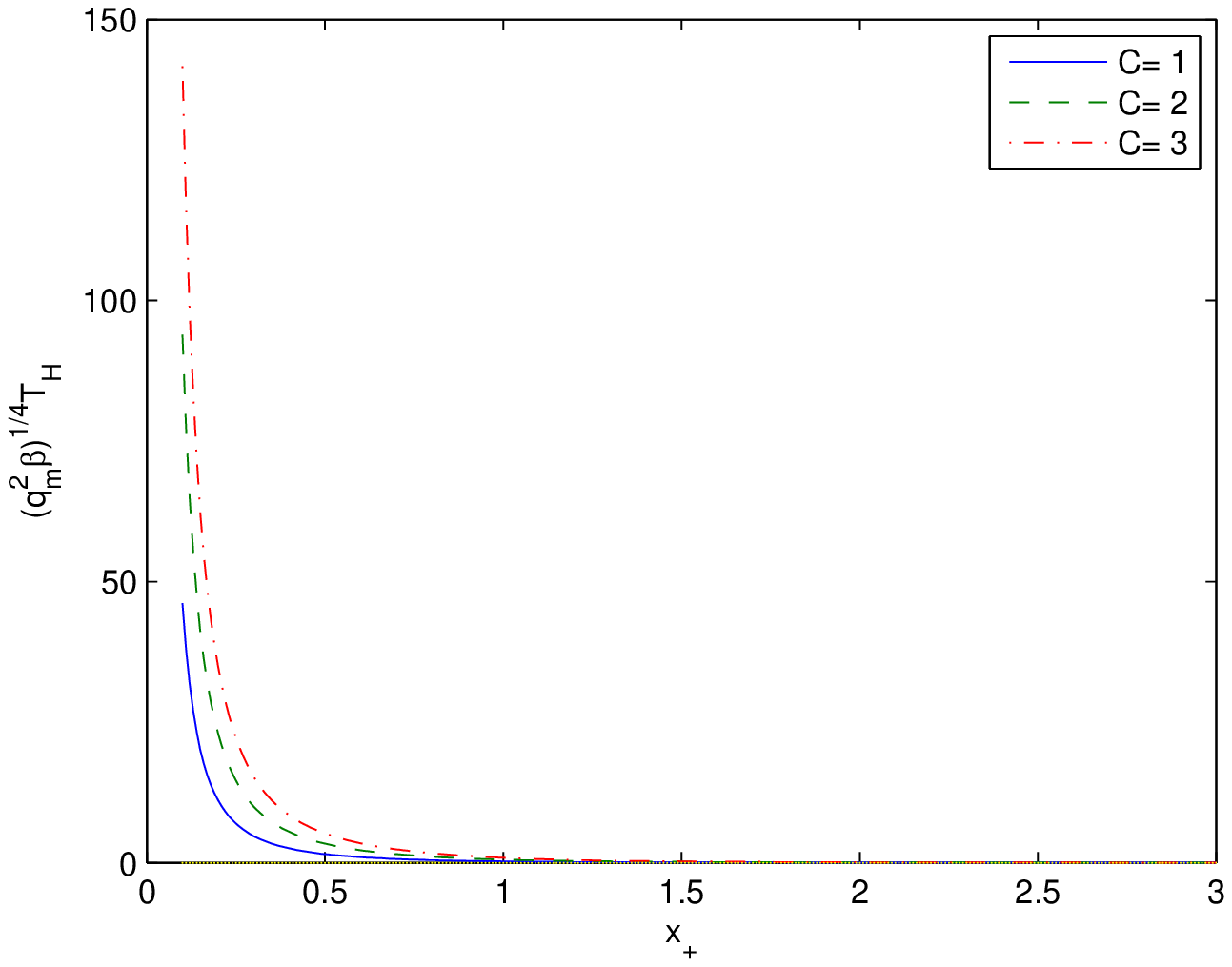}
\caption{\label{fig.3}The plot of the function $T_H\sqrt{q_m}\beta^{1/4}$ vs $x_+$.  The solid curve is for $C=1$, the dashed curve corresponds to $C=2$, and the dashed-doted curve corresponds to $C=4$.}
\end{figure}
\begin{figure}[h]
\centering
\includegraphics[height=2.in,width=2.in]{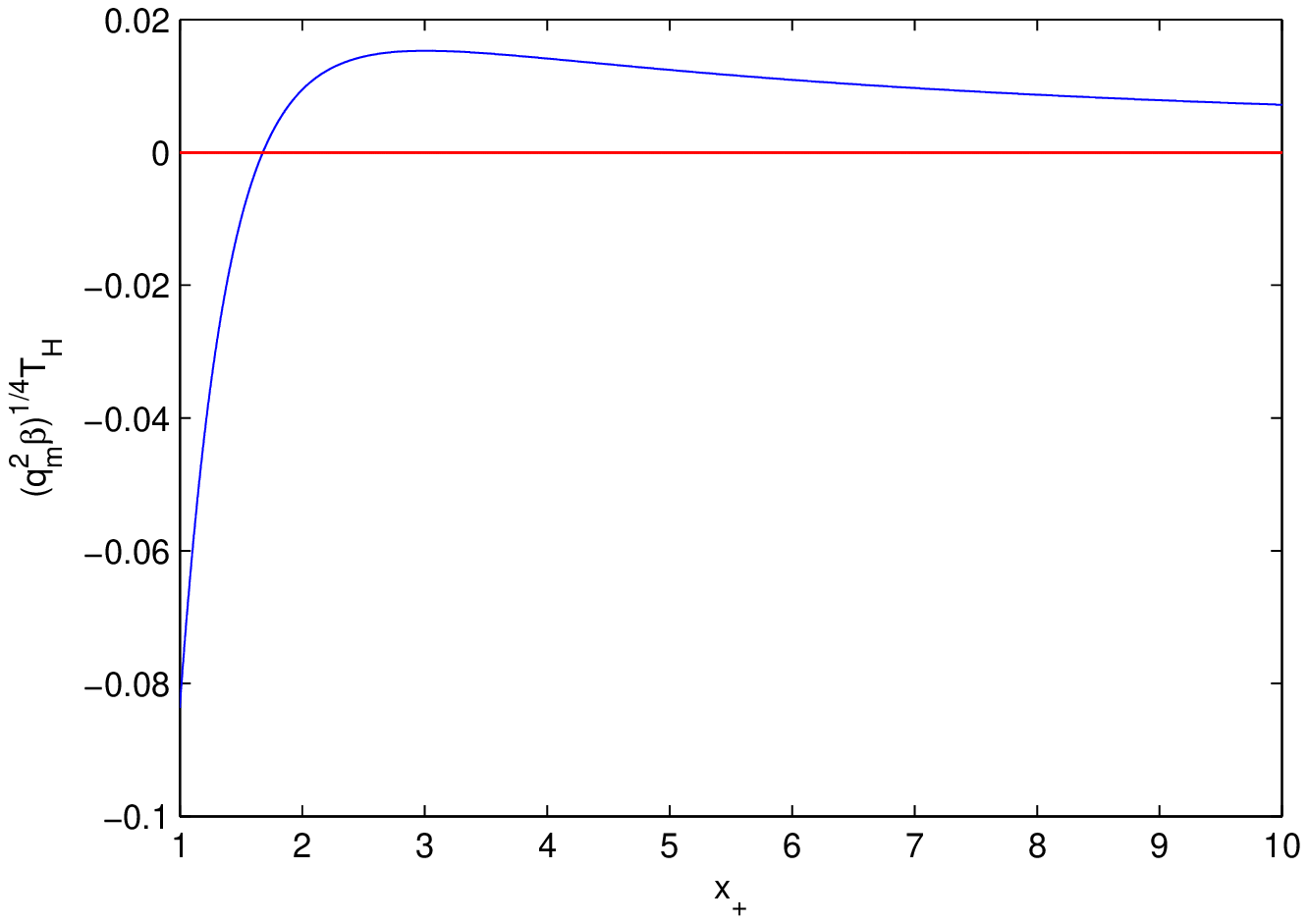}
\caption{\label{fig.4}The plot of the function $T_H\sqrt{q_m}\beta^{1/4}$ vs $x_+$ for $C=0$ ($m_0=0$). }
\end{figure}
According to Fig. 3 the temperature is positive everywhere for the case $C\neq 0$ ($m_0\neq 0$).
Figure 4 shows that the Hawking temperature for $C=0$ ($m_0=0$) is positive for $x_+>1.679 $ and is zero at $x_+\approx 1.679$ . The BH is unstable when the temperature is negative. From Eq. (30) we obtain the value of $Gq_m/\sqrt{\beta}=3.173$ corresponding to $x_+=1.679$.
By studying the signs of the heat capacity and the Helmholtz free energy, we can observe the different stability phases of the BH \cite{Page}. Making use of the  Hawking entropy of the BH $S=\mbox{Area}/(4G)=\pi r_+^2/G=\pi x_+^2q_m\sqrt{\beta}/G$ we find the heat capacity
\begin{equation}
C_q=T_H\left(\frac{\partial S}{\partial T_H}\right)_q=\frac{T_H\partial S/\partial x_+}{\partial T_H/\partial x_+}=\frac{2\pi q_m\sqrt{\beta}x_+T_H}{G\partial T_H/\partial x_+}.
\label{32}
\end{equation}
According to Eq. (32) the heat capacity possesses a singularity when the Hawking temperature has an extremum ($\partial T_H/\partial x_+=0$). The plots of the heat capacity versus the variable $x_+$ for different parameters $C$ are depicted in Figs. 5. 6, and 7.
\begin{figure}[h]
\centering
\includegraphics[height=2.in,width=2.in]{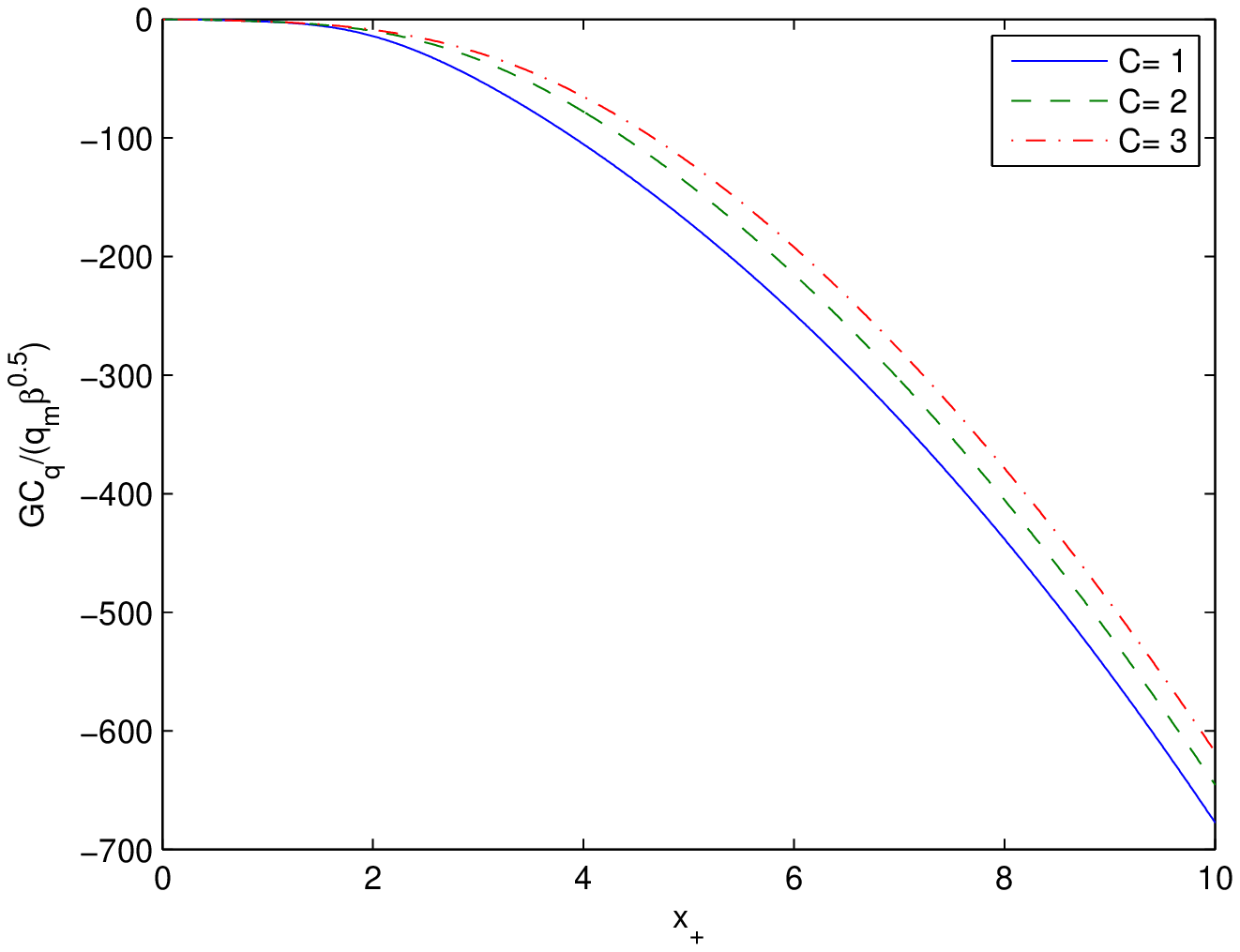}
\caption{\label{fig.5}The plot of the function $GC_q/(q_m^2\beta)^{1/2}$ vs $x_+$. The solid curve is for $C=1$, the dashed curve corresponds to $C=2$, and the dashed-doted curve corresponds to $C=3$.}
\end{figure}
\begin{figure}[h]
\centering
\includegraphics[height=1.8in,width=1.8in]{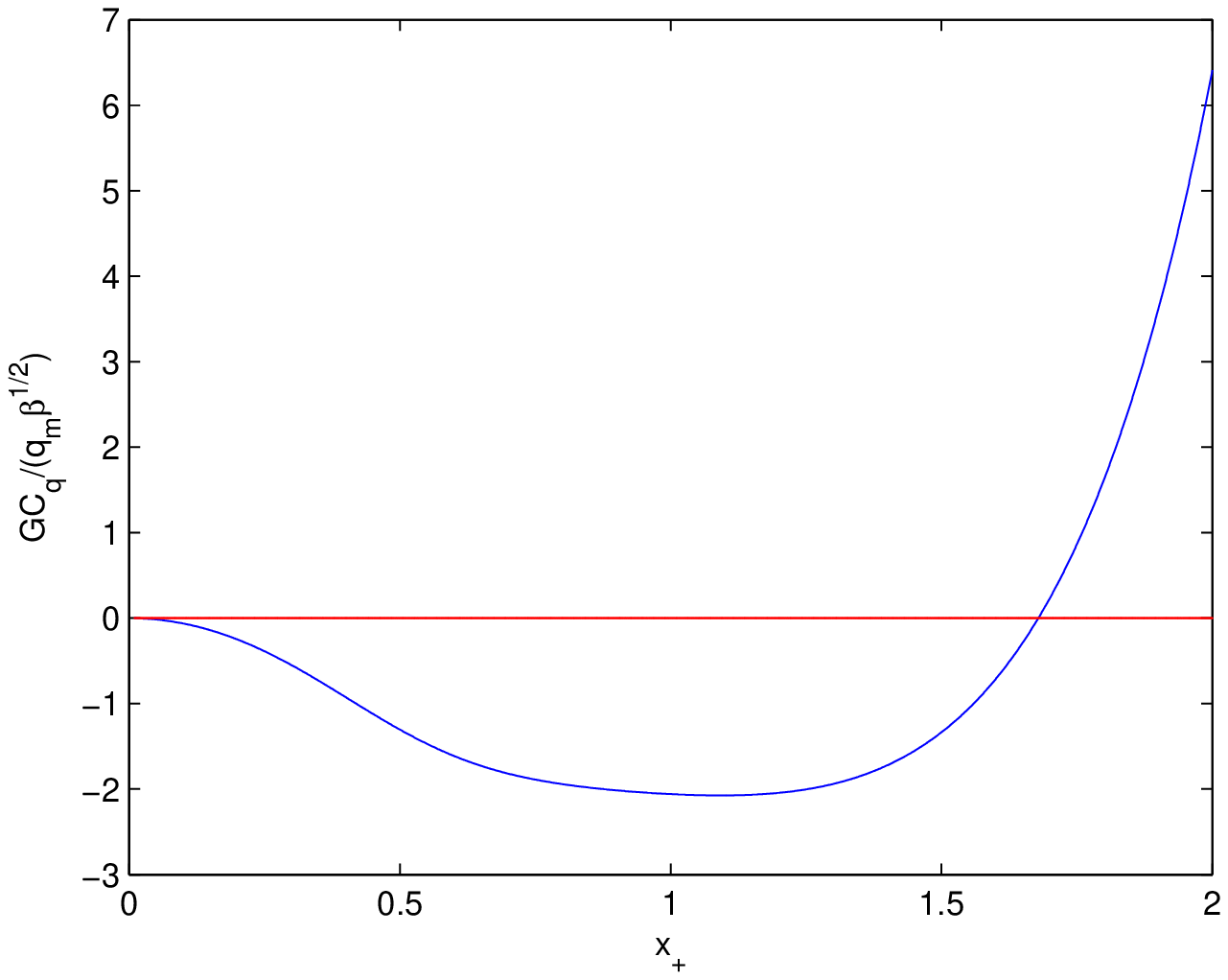}
\caption{\label{fig.6}The plot of the function $GC_q/(q_m^2\beta)^{1/2}$ vs $x_+$ for $C=0$.}
\end{figure}
\begin{figure}[h]
\centering
\includegraphics[height=1.8in,width=1.8in]{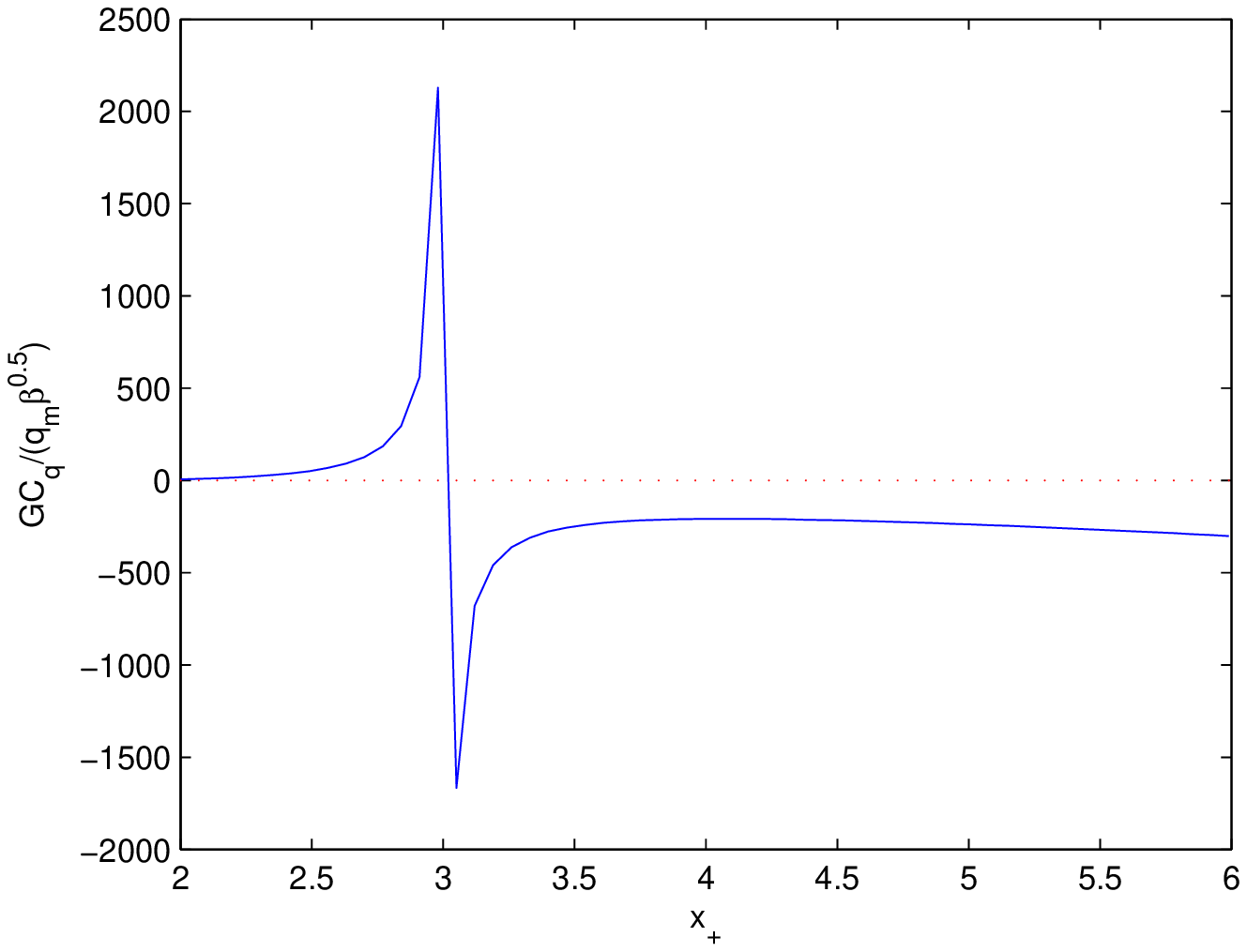}
\caption{\label{fig.7}The plot of the function $GC_q/(q_m^2\beta)^{1/2}$ vs $x_+$ for $C=0$.}
\end{figure}
Figure 5 shows the Schwarzschild behaviour of the heat capacity for $C\neq 0$ ($m_0\neq 0$), i.e. it is negative at $\partial T_H/\partial x_+<0$. As a result, the BHs are unstable for the case $m_0\neq0$. In accordance with Fig. 6  the BH is unstable at $1.679>x>0$ because the heat capacity is negative.
Figure 7 shows a singularity in the heat capacity at the point $x\approx 3$ where the second-order phase transition occurs. When $m_0=0$ the heat capacity is positive at the range $3>x>1.679$ and the BH is stable.

To complete the analysis of phase transitions we consider the Helmholtz free energy which is given by
\begin{equation}
F=m-T_HS.
\label{33}
\end{equation}
The mass of the BH $m$ plays the role of the internal energy, and the Hawking entropy is $S=\pi r_+^2/G$. Making use of Eqs. (22). (31) and (33) we obtain
\begin{equation}
\frac{GF}{\sqrt{q_m}\beta^{1/4}}=B\left(C+\frac{\pi}{4\sqrt{2}}\right)-\frac{x_+}{4}+
\frac{\sqrt{2}x_+^2(x_+-2C)}{(x_+^4+1)D}.
\label{34}
\end{equation}
Substituting $B=q_mG/\sqrt{\beta}$ from Eq. (30) into (34) we find
\begin{equation}
\frac{GF}{\sqrt{q_m}\beta^{1/4}}=\left(C+\frac{\pi}{4\sqrt{2}}\right)\frac{4\sqrt{2}(x_+-2C)}{D}-\frac{x_+}{4}+
\frac{\sqrt{2}x_+^2(x_+-2C)}{(x_+^4+1)D}\biggr).
\label{35}
\end{equation}
The plots of the unitless reduced free energy $GF/(\sqrt{q_m}\beta^{1/4})$ vs. $x_+$ are depicted in Figs. 8 and 9.
\begin{figure}[h]
\centering
\includegraphics[height=1.8in,width=1.8in]{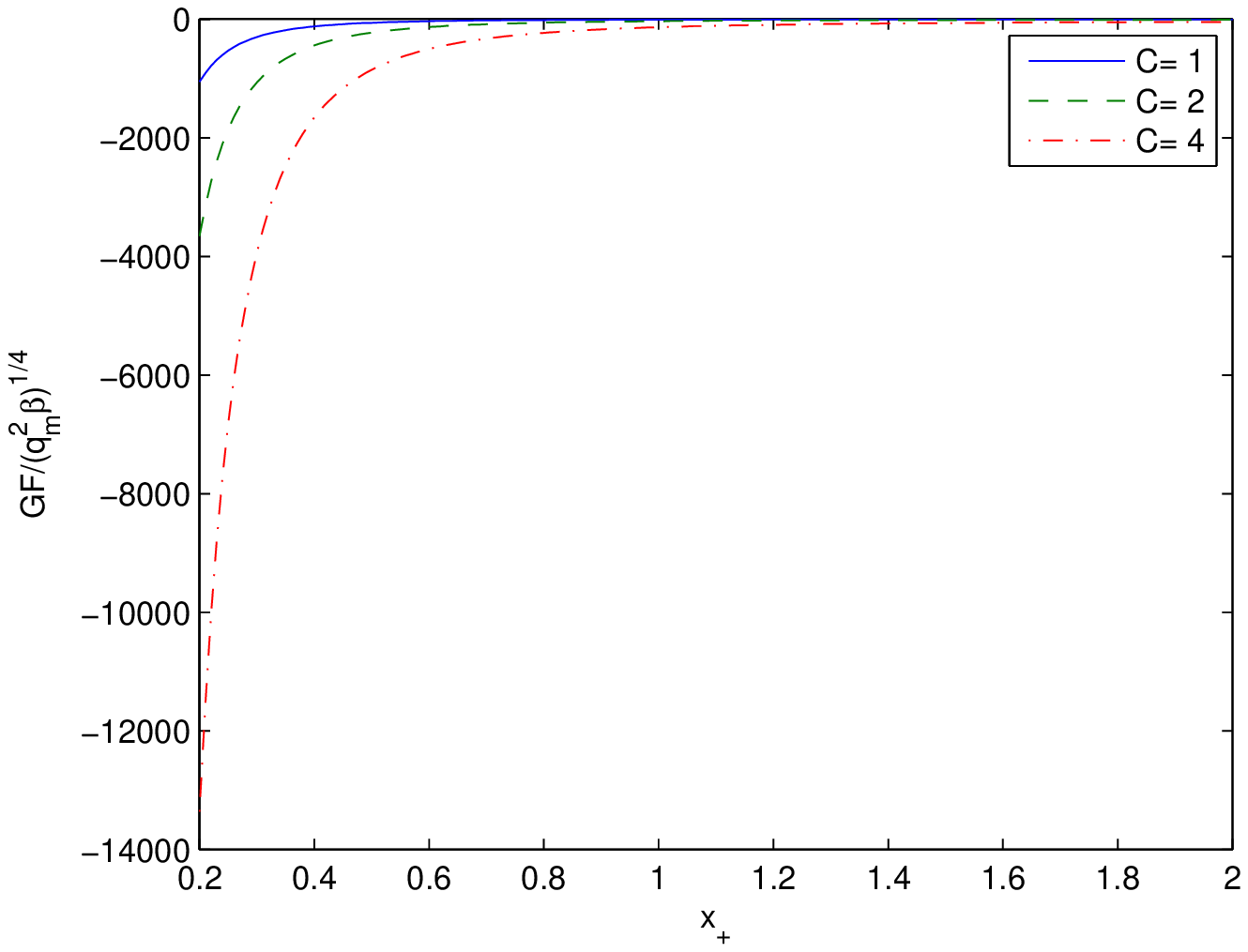}
\caption{\label{fig.8}The plot of the function $GF/(\sqrt{q_m}\beta^{1/4})$ vs. $x_+$. The dashed curve corresponds to $C=2$, the solid curve is for $C=1$, and the dashed-doted curve corresponds to $C=4$.}
\end{figure}
\begin{figure}[h]
\centering
\includegraphics[height=1.8in,width=1.8in]{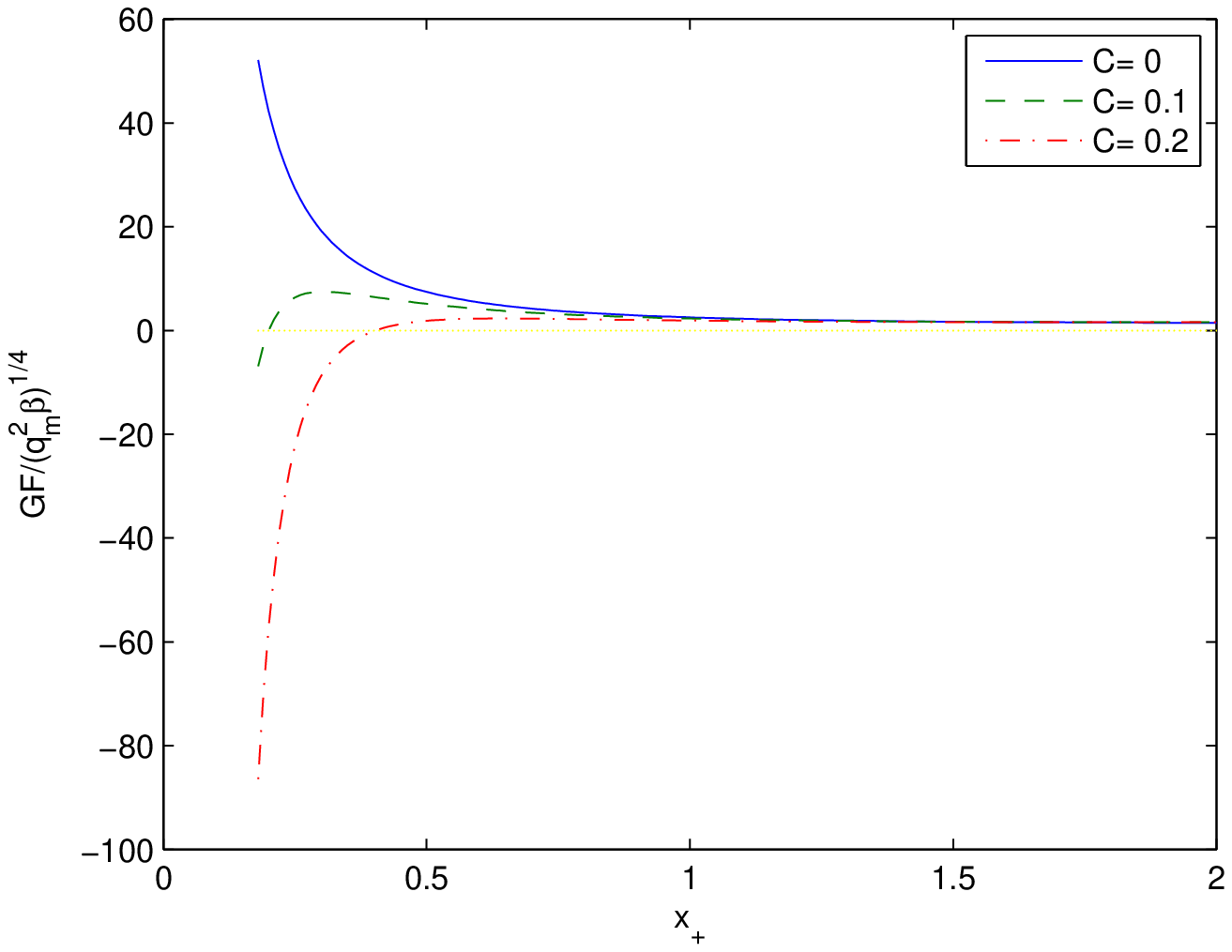}
\caption{\label{fig.9}The plot of the function $GF/(\sqrt{q_m}\beta^{1/4}$ vs. $x_+)$. The dashed curve corresponds to $C=0.1$, the solid curve is for $C=0$, and the dashed-doted curve corresponds to $C=0.2$.}
\end{figure}
The BHs with $F > 0$, $C_q < 0$ are unstable and the BHs with $F < 0$, $C_q > 0$ are stable. In accordance with Figs. 5-9, there are other phases with $F > 0$, $C_q > 0$  and $F <0$, $C_q < 0$. In the case  $F < 0$, $C_q < 0$ the BHs are less energetic than the pure radiation and, as a result, BHs do not decay through tunneling. For large masses of BHs ($C>1$) this phase holds. Because the heat capacities are negative, the BH temperature decreases when the mass of BH increases. Such phases are also realised in another model \cite{Jarillo}.

 \section{Conclusion}

The correspondence principle of the NED model holds as for weak fields our model is converted into Maxwell's electrodynamics. It was demonstrated that at $\beta\textbf{B}^2\leq 1/3$ ($\textbf{E}=0$) the principles of causality and unitarity take place. In this model the singularity of the electric field at the center of charges is absent and the maximum electric field in the origin is $E(0)=1/\sqrt{\beta}$.
The dyonic and magnetic BHs in GR were studied. It was shown that in the self-dual case ($q_e=q_m$) the corrections to Coulomb's law and RN solutions are absent. The Ricci scalar does not have the singularity and as $r\rightarrow \infty$ space-time becomes flat.

The thermodynamics and the thermal stability of magnetized BHs were investigated. The Hawking temperature, the heat capacity and the Helmholtz free energy of BHs were calculated. It was demonstrated that the heat capacity diverges at some event radii $r_+$ ($x_+$) for the case when the total BH mass is the magnetic mass and the phase transitions of the second-order occurs.
We shown that there is a new stability region of BH solutions when the heat capacity and the free energy are negative. In this case BHs are less energetic than the pure radiation and BHs do not decay via tunneling.

\end{document}